\documentclass[twocolumn]{webofc}

\usepackage[varg]{txfonts}   
\begin{document}
\title{Radio detection of cosmic rays with the Auger Engineering Radio Array}
%
%

\author{\firstname{Tim} \lastname{Huege}\inst{1,3}\fnsep\thanks{\email{tim.huege@kit.edu}} 
        on behalf of \lastname{the Pierre Auger Collaboration}\inst{2}\fnsep\thanks{Full author 
        list at http://www.auger.org/archive/authors\_2018\_10.html}
}

\institute{Karlsruhe Institute of Technology (KIT), Institute for Nuclear Physics, Postfach 3640, 76021 Karlsruhe, Germany 
\and
           Observatorio Pierre Auger, Av. San Mart\'in Norte 304, 5613 Malarg\"ue, Argentina
\and
           also at: Astrophysical Institute, Vrije Universiteit Brussel, Pleinlaan 2, 1050 Brussels, Belgium
          }

\abstract{%
The Auger Engineering Radio Array (AERA) complements the Pierre Auger Observatory with 150 radio-antenna stations measuring in the
frequency range from 30 to 80~MHz. With an instrumented area of 17 km$^2$, the array constitutes the largest cosmic-ray radio detector
built to date, allowing us to do multi-hybrid measurements of cosmic rays in the energy range of 10$^{17}$~eV up to several 10$^{18}$~eV.
We give an overview of AERA results and discuss the significance of radio detection for the validation of the energy scale of
cosmic-ray detectors as well as for mass-composition measurements.
}
\maketitle
%


\section{Introduction} \label{intro}

The Pierre Auger Observatory \cite{AugerNIM2014} exploits hybrid detection of extensive air 
showers measured at the same time with multiple detection techniques 
for optimum determination of the energy and mass of cosmic ray 
primaries. Following this spirit, its baseline surface detector (SD) 
and fluorescence detectors (FD)  have been complemented by an 
array of radio detectors. This Auger Engineering Radio Array (AERA) 
was deployed in the region where many enhancements, focused in 
particular at the detection of air showers with energies in the EeV 
energy regime, have been set up.

In this article, we briefly review the characteristics and capabilities 
of the radio detectors of the Pierre Auger Observatory and discuss how 
they contribute valuable information in particular for the 
determination of the energy scale of cosmic rays as well as mass 
composition studies.


\section{The Auger Engineering Radio Array} \label{sec:aera}

When work on AERA began in 2008, the nature 
and characteristics of the radio emission from extensive air showers
were still largely unclear --- unlike 
today, where the emission physics and signal characteristics are 
very-well understood \cite{HuegePLREP,SchroederReview}. Furthermore, 
many questions regarding the best strategy for the measurement of these 
pulsed radio signals were still open. The goal of AERA, illustrated by the 
explicit mentioning of ``engineering'' in its name, was to explore different options 
for implementing a radio detector that can be scaled to areas 
significantly larger than a km$^2$.

These efforts resulted in the setup of a 17~km$^2$ array of 
radio detectors. The individual detector stations are \emph{autonomous} 
and can thus be freely spaced. Many of the experimental challenges, for 
example related to power supply, communications and data acquisition, were 
indeed related to this autonomous design, which is a key and unique feature of 
AERA that differentiates it from purely cabled setups for radio 
detection of cosmic rays.

\begin{figure*}
\centering
\includegraphics[width=0.65\textwidth]{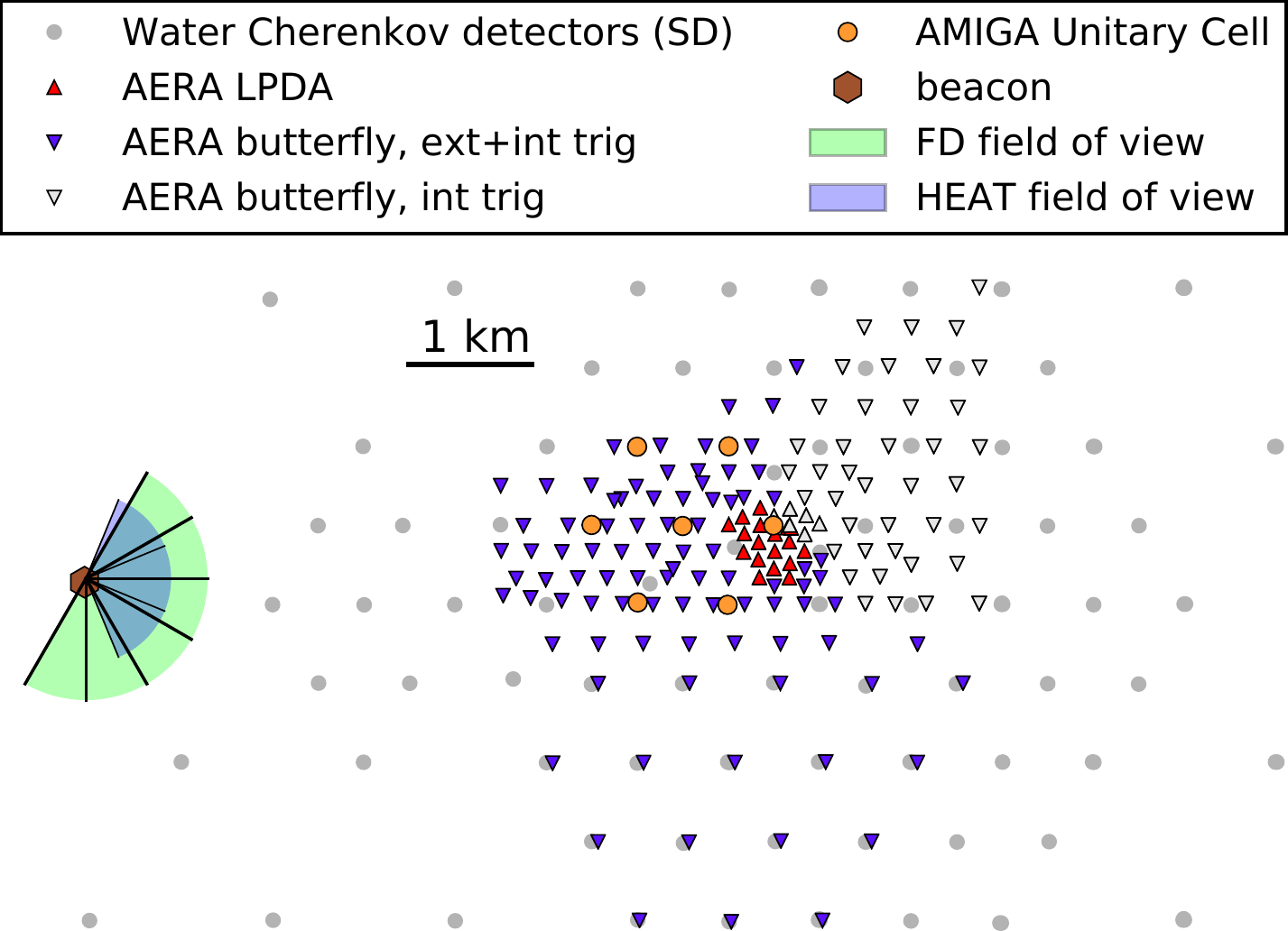}
\caption{Map of the Auger Engineering Radio Array. 150 radio-detector 
stations cover a total area of 17~km$^2$.
Further explanations are given in the text.}
\label{map}       
\end{figure*}

The layout of the array is depicted in Figure \ref{map}. The radio 
antennas are co-located with the 750~m surface detector array, the 
Coihueco and HEAT fluorescence telescopes, and the AMIGA 
underground muon detectors. This combination of detectors at the same 
location maximizes the information 
gathered on each individual air shower by multi-hybrid detection.

The antennas have been set up on a graded array, with grid constants of 
144, 250, 375 and 750 meters, probing different energy and zenith angle 
ranges. Two different types of antennas, Logarithmic Periodic Dipole 
Antennas (LPDAs) \cite{AERAAntennaPaper2012,AERALPDA2017} and 
``butterfly antennas'' \cite{AERAAntennaPaper2012} are in use. Both 
measure in the frequency band from 30 to 80 MHz. Furthermore, 
two different types of digital electronics, both of them capable of 
``internal'' triggering on radio signals and one of them, in addition, allowing data 
buffering for up to 8 seconds and thus exploitation of external 
triggering by the SD and FD, have been developed and deployed. A 
reference ``beacon'' transmitter is used to achieve time 
synchronization of the autonomous detectors within two nanoseconds 
\cite{AERAAirplane}.

AERA started taking data in 2011 and has been finalized in its 
current form in 2015. Its current dataset encompasses more than 10,000 
extensive air showers measured with both the SD and the radio 
antennas. In the following, we give a concise overview of the most important 
results derived with AERA up to now.



\section{Determination of the cosmic-ray energy} \label{sec:energyscale}

One of the key parameters to characterize cosmic-ray primaries is their 
energy. With AERA, we have demonstrated that this energy can be 
determined with a resolution of 17\% or better \cite{AERAEnergyPRD}. Many different 
approaches exist to determine the cosmic-ray energy from radio 
measurements \cite{HuegePLREP}. In AERA, we determine the \emph{energy 
fluence}, i.e. the energy deposited in the form of radio signals per 
unit area, at the locations of our radio detector stations. Exploiting 
our knowledge of the radio-emission pattern, we can then 
interpolate and finally integrate the energy fluence to determine the 
total energy deposited on the ground in the form of radio signals, the 
\emph{radiation energy}.

\begin{figure}[htb]
\centering
\includegraphics[width=0.49\textwidth]{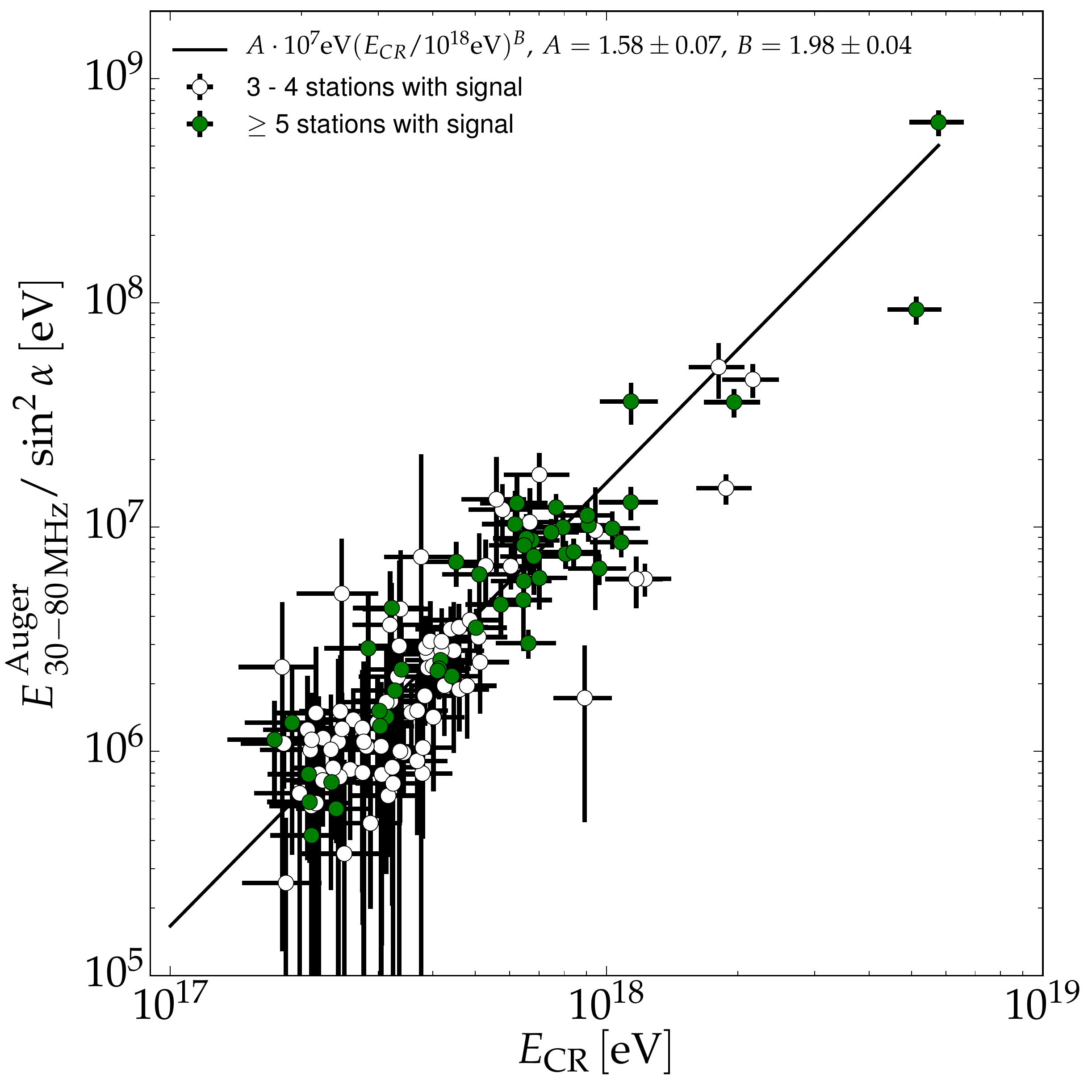}
\caption{After correction for the angle of 
incidence with respect to the geomagnetic field $\alpha$,
the energy deposited on the ground in the form of radio 
signals, the \emph{radiation energy}, scales quadratically with the 
energy of the primary cosmic ray.
With AERA, an energy resolution of 17\% 
has so far been demonstrated \cite{AERAEnergyPRL} for events detected 
in at least 5 radio-detector stations with signal.}
\label{energy}       
\end{figure}

After correction for the orientation with regard to the geomagnetic 
field, this radiation energy is closely correlated with the energy in the 
electromagnetic cascade of the air shower, following a quadratic scaling 
because of the coherent nature of the radio emission. This is shown in 
Figure \ref{energy}. As the radio emission undergoes no significant 
absorption or scattering in the atmosphere and constitutes a 
calorimetric quantity, it is very well suited for cross-calibration of 
cosmic-ray detectors via radio measurements \cite{AERAEnergyPRL}. Comparing the measured 
radiation energy to the one predicted by first-principle Monte Carlo 
simulations \cite{Gottowik:2017wio} also provides the opportunity for an independent cross 
check and validation of the energy scale of the Pierre Auger 
Observatory \cite{EnergyScaleICRC2013}, which is based on measurements of the fluorescence 
detectors.

\section{Mass composition sensitivity} \label{sec:masss}

The depth of maximum $X_{\mathrm{max}}$ of an extensive air shower is a 
mass-sensitive parameter that is directly accessible with fluorescence 
measurements. Radio signals from air showers also have sensitivity to 
$X_{\mathrm{max}}$ due to the 
forward-beamed nature of the emission as well as its enhancement on a
Cherenkov ring introduced by the non-unity refractive index of the 
atmosphere \cite{HuegePLREP}. Deeper showers thus produce a steeper signal falloff with 
lateral distance from the impact point than shallower showers. In 
addition, pulse shape and polarization information can be exploited.

\begin{figure}[t]
\centering
\includegraphics[width=0.49\textwidth,trim=0cm 0cm 10cm 0cm,clip=true]{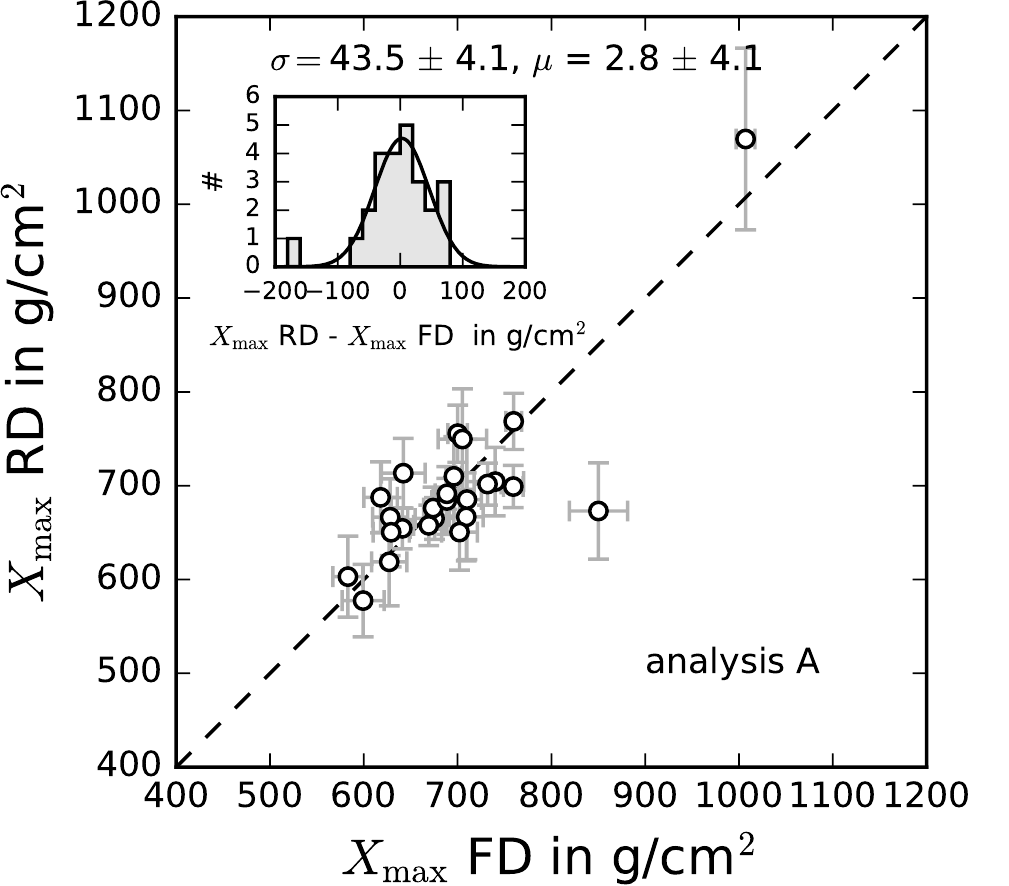}
\caption{The depth of shower maximum $X_{\mathrm{max}}$ as determined 
with AERA correlates well with the $X_{\mathrm{max}}$ value determined 
from the Auger fluorescence detectors \cite{HoltICRC2017}. The combined 
FD-radio resolution is approximately 45 g/cm$^2$, and the radio-only resolution is of order 35-40 
g/cm$^2$.}
\label{xmax}       
\end{figure}

We have demonstrated that by comparison of the radio signals of 
measured air showers with per-event Monte Carlo simulations, 
$X_{\mathrm{max}}$ can be determined and is in agreement with the 
values measured for the same showers with the fluorescence detectors 
\cite{HoltICRC2017}, 
see Figure \ref{xmax}. 
The combined FD-radio resolution is approximately 45 g/cm$^2$, i.e. the radio-only 
resolution is of order 35-40 g/cm$^2$. This is likely limited by the 
sparsity of the radio array, as measurements with much denser radio arrays have 
yielded $X_{\mathrm{max}}$ resolutions below 20 g/cm$^2$ 
\cite{LOFARXmaxMethod2014}.

\begin{figure}[htb]
\centering
\includegraphics[width=0.49\textwidth]{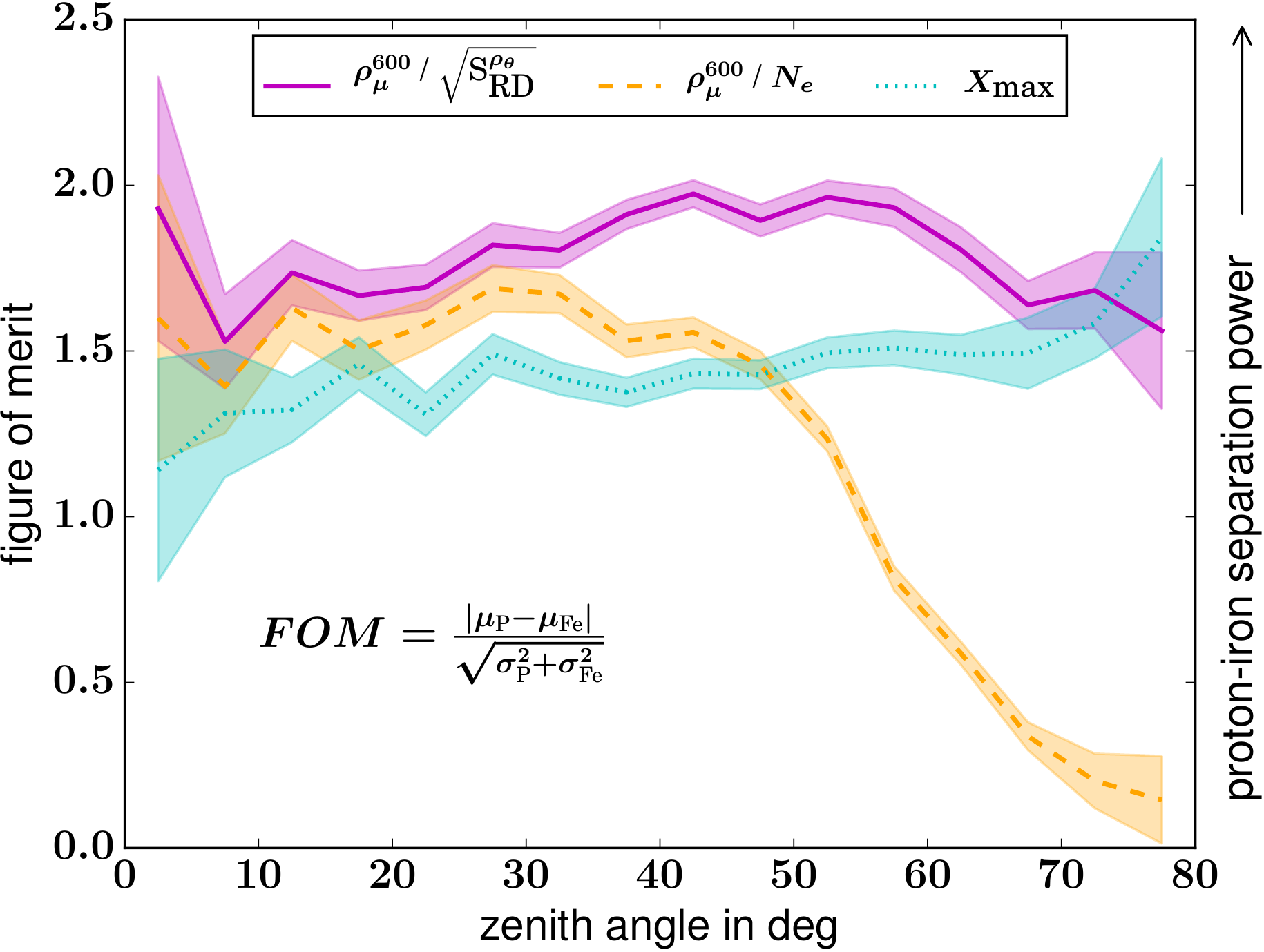}
\caption{A simulation study of the separation power between proton- and 
iron-induced air showers shows that combined measurements of the radiation 
energy, $S_{\mathrm{RD}}^{\rho_\theta}$, and the muon content at a 
lateral distance of 600 meters, $\rho_{\mu}^{600}$, provide
very good mass composition sensitivity at all zenith angles 
\cite{HoltARENA2018}. In contrast, combined measurements of the electron number 
$N_{e}$ and the muon content lose separation power towards higher zenith 
angles, as the electromagnetic cascade dies out before the shower 
reaches the ground. Detector effects are not included in this study.}
\label{muons}       
\end{figure}

Another option for mass composition studies with radio antennas lies in 
their pure measurement of the energy in the electromagnetic component 
of an air shower. Combined with a measurement of the muon content of 
air showers, for example by the AMIGA underground muon detectors, this 
allows mass-composition studies through a completely different method. 
This approach has particular potential for inclined air showers, for 
which a classical measurement of the electron number at the ground is 
no longer feasible (the electromagnetic cascade dies out before 
reaching the ground due to the large atmospheric mass overburden). As 
radio signals are not significantly attenuated in the atmosphere, radio 
detection at these geometries allows the determination of the energy in the electromagnetic 
cascade, while particle detectors such as those of the SD will provide 
a clean measurement of the muonic cascade. The sensitivity for 
mass-composition measurements arising from this complementarity is illustrated in 
Figure \ref{muons}.

\section{Radio measurements of inclined air showers} \label{sec:has}

The radio emission from extensive air showers is strongly 
forward-beamed. For showers with near-vertical incidence, for which the 
shower maximum is located at a few kilometers above sea level, the 
emission thus illuminates a rather small area on the ground, of order 
hundreds of meters in diameter \cite{HuegeUHECR2014}. The area does not grow significantly 
with the energy of the primary particle because of the very steep 
lateral falloff of the radio signal and the even deeper depth of 
maximum for higher particle energies. As a consequence, radio detector 
arrays aiming to measure near-vertical air showers need a dense antenna 
spacing with a grid constant of at most a few hundred meters. Scaling 
such arrays significantly beyond the size of AERA is thus not cost-effective. 
For near-vertical showers, the energy reach is therefore limited to 
below 10$^{19}$~eV, simply because the 
instrumentable area is limited.

\begin{figure*}[h!tb]
\centering
\includegraphics[width=0.8\textwidth]{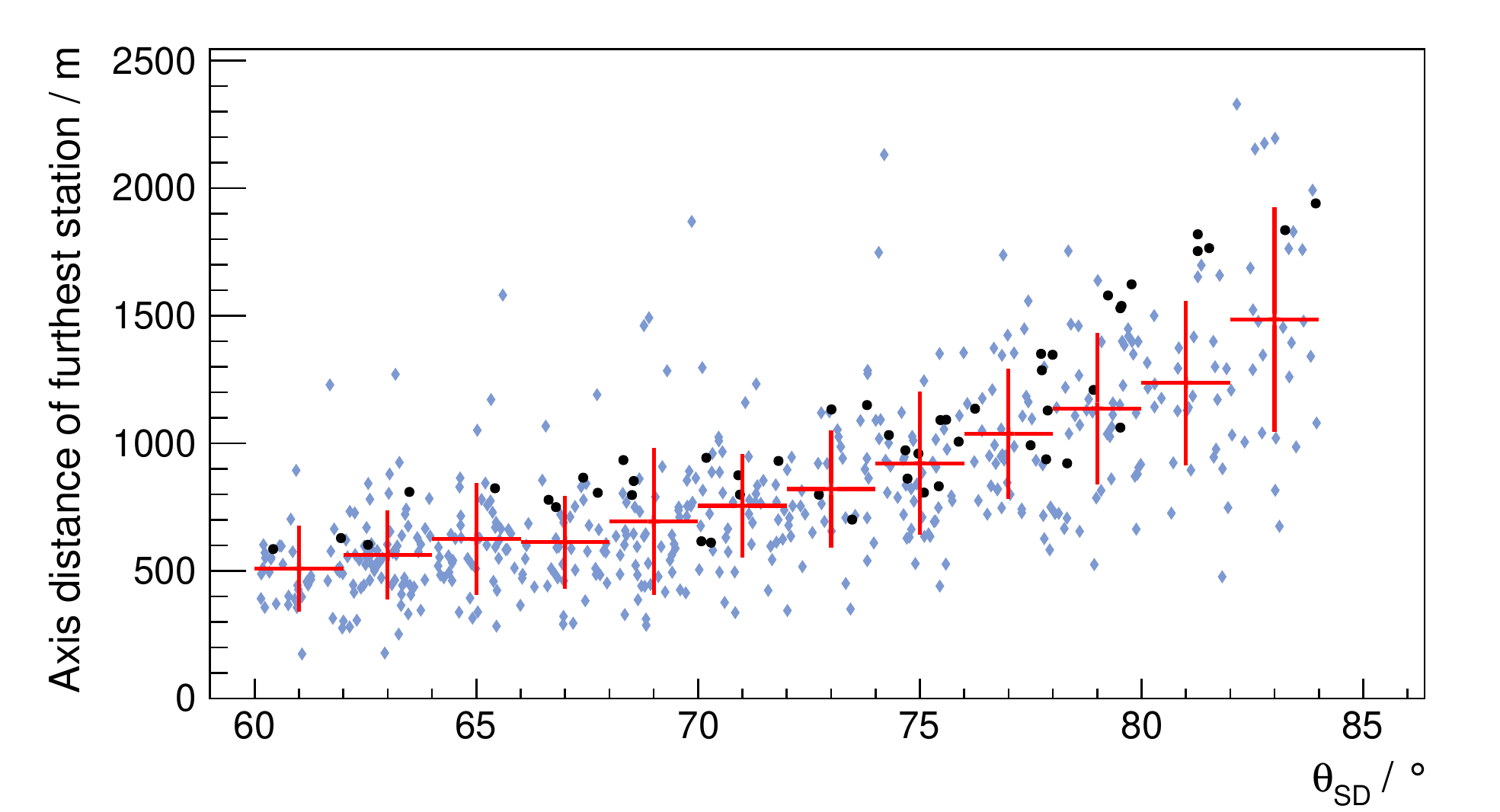}
\caption{With increasing zenith angle, the farthest distance from the shower 
axis at which a detectable radio signal was measured by AERA grows 
significantly. At the highest zenith angles, detectable radio signals 
illuminate areas larger than 100~km$^2$.}
\label{has}       
\end{figure*}

In contrast, for inclined air showers, due to the vastly increased atmospheric 
mass overburden, the shower maximum is typically dozens of kilometers 
away from a ground-based antenna array. Consequently the emission is 
bound to be distributed over a much larger area \cite{HuegeUHECR2014}.
This has recently been confirmed experimentally 
with AERA data \cite{Aab:2018ytv} as presented in Figure \ref{has}: The distance of the 
radio detector station farthest from the air-shower axis which still 
measured a clear radio pulse increases significantly with the air-shower 
zenith angle. At the highest zenith angles, the lateral extent of the 
radio-emission footprint can exceed two kilometers in the shower plane. 
Projection to the ground then yields areas of more than $2$~km$\, \times\, 
2$~km$\, \times \sec(83^{\circ}) \times\,\pi = 103$~km$^{2}$ illuminated with measurable radio signals.
This allows radio detection of inclined air showers with very 
sparse radio arrays. In fact, we have demonstrated that a grid constant 
of 1.5~km, the same as that of the Auger SD, is sufficient for the 
detection of inclined air showers \cite{Aab:2018ytv}.
For inclined air showers, 
radio detection then provides a pure measurement of the electromagnetic 
component of the air shower, while the SD measurements provide the pure 
muon content. As described above, this opens the possibility for 
mass-composition studies of inclined air showers.

Motivated by these findings, we will equip each water-Cherenkov 
detector of the SD with one radio antenna as part of the ongoing 
AugerPrime upgrade \cite{Aab:2016vlz}. The details of this endeavor
are described in a dedicated article \cite{HorandelUHECR2018}.


\section{Conclusions} \label{sec:conclusions}

The Auger Engineering Radio Array, an array of 150 autonomous radio detector stations 
covering a total area of 17~km$^2$, measures extensive air showers in 
the energy range of 10$^{17}$~eV to several 10$^{18}$~eV via 
their radio emission in the 30 to 80~MHz band. AERA explored many 
different approaches for the experimental design as well as several analysis 
strategies, and has made important contributions to leading the radio 
detection technique from pioneering prototypes to maturity.

In particular, we have demonstrated that the energy deposited in the 
form of radio waves on the ground, the \emph{radiation energy}, can be 
measured accurately and is a reliable estimator for the energy in the 
electromagnetic cascade of an extensive air shower. This quantity has a 
well-defined physical meaning and can be measured and compared among 
radio detectors worldwide. As such, it is a very useful means to 
cross-calibrate the energy scale of different cosmic-ray experiments, 
and can even be used to validate the energy scale on the basis of first 
principle calculations.

Mass composition sensitivity is also present in radio signals. We 
have successfully reconstructed the depth of shower maximum of air 
showers from radio measurements with AERA, and they are in agreement 
with measurements from the Auger fluorescence detector. Another way to determine the mass of 
primary cosmic rays with the help of radio measurements lies in  
combining radio detection with muon measurements, as are provided for 
near-vertical geometries by the Auger AMIGA detectors and for inclined air 
showers by the Auger surface detector.

Finally, we have established experimentally that inclined air showers 
illuminate areas of dozens or even more than a hundred km$^2$ with measurable radio signals. 
Hence, arrays with grid constants of a kilometer or larger can measure 
inclined air showers with reasonable antenna multiplicity. This allows 
radio measurements up to ultra-high energies, and yields very 
complementary information to that provided by the water-Cherenkov 
detectors of the SD. Consequently, as part of the AugerPrime upgrade, 
we will deploy a radio antenna on top of each water-Cherenkov detector 
to extend the composition-sensitive measurements of AugerPrime to inclined
air showers. At the same time, we will continue to operate AERA, which 
is focused at energies around the EeV scale.



\begin{thebibliography}{17}

\bibitem{AugerNIM2014}
A.~{Aab} {et al.} (Pierre Auger Collaboration),
  Nucl. Instrum. Meth. A \textbf{798}, 172 (2015)

\bibitem{HuegePLREP}
T.~Huege, Physics Reports \textbf{620}, 1  (2016)

\bibitem{SchroederReview}
F.~G. Schr\"oder, Progress in Particle and Nuclear Physics \textbf{93}, 1
  (2017)

\bibitem{AERAAntennaPaper2012}
P.~{Abreu} {et al.} (Pierre Auger Collaboration), JINST \textbf{7}, P10011
  (2012)

\bibitem{AERALPDA2017}
A.~{Aab} {et al.} (Pierre Auger Collaboration), JINST \textbf{12}, T10005 (2017)

\bibitem{AERAAirplane}
A.~Aab {et al.} (Pierre Auger Collaboration),
  JINST \textbf{11}, P01018 (2016)

\bibitem{AERAEnergyPRD}
A.~Aab {et al.} (Pierre Auger Collaboration), Phys.
  Rev. D \textbf{93}, 122005 (2016)

\bibitem{AERAEnergyPRL}
A.~Aab {et al.} (Pierre Auger Collaboration), Phys.
  Rev. Lett. \textbf{116}, 241101 (2016)

\bibitem{Gottowik:2017wio}
M.~Gottowik, C.~Glaser, T.~Huege, J.~Rautenberg, Astropart. Phys. \textbf{103},
  87 (2018)

\bibitem{EnergyScaleICRC2013}
V.~{Verzi}, {for the Pierre Auger Collaboration}, Proc. 33rd ICRC, Rio de
  Janeiro, Brazil, id 0928 (2013)

\bibitem{HoltICRC2017}
E.~M. {Holt} {for the Pierre Auger Collaboration}, Proc. 35th ICRC,
  Busan, Korea, PoS(ICRC2017)492

\bibitem{LOFARXmaxMethod2014}
S.~{Buitink} {et al.} (LOFAR CR KSP), Phys.\ Rev.\ D
  \textbf{90}, 082003 (2014)

\bibitem{HoltARENA2018}
E.~M. {Holt} {for the Pierre Auger Collaboration}, Proc. of the 2018 
ARENA conference, EPJ WoC in press (2019)

\bibitem{HuegeUHECR2014}
T.~{Huege}, A.~{Haungs}, Proc. of the UHECR2014 conference, Springdale, 
USA, JPS Conf. Proc. \textbf{09}, 010018 (2016)

\bibitem{Aab:2018ytv}
A.~Aab {et al.} (Pierre Auger Collaboration), JCAP \textbf{10}, 026 (2018)

\bibitem{Aab:2016vlz}
A.~Aab {et~al.} (Pierre Auger Collaboration), ArXiv e-prints, \texttt{1604.03637} (2016)

\bibitem{HorandelUHECR2018}
J.~{H\"orandel} {for the Pierre Auger Collaboration}, these 
proceedings

\end{thebibliography}

\end{document}